\documentclass[12pt]{iopart}
\usepackage{graphicx}

\usepackage[usenames]{color}

\begin{document}

\title{Stability of trapped degenerate dipolar Bose and Fermi gases}


\author{S. K. Adhikari 
} 
\address{Instituto de F\'{\i}sica Te\'orica,
UNESP - Universidade Estadual Paulista,\\ 01.140-070 S\~ao Paulo, S\~ao
Paulo, Brazil}

\begin{abstract} Trapped degenerate dipolar Bose and Fermi gases of cylindrical symmetry
with the polarization vector along the symmetry axis  
are only stable for the strength of
dipolar interaction below a critical value. In the case of bosons, the stability of such a dipolar 
Bose-Einstein condensate (BEC) 
is investigated for different strengths of contact and 
dipolar interactions using  variational   approximation and numerical 
solution of a mean-field model. In the disk shape, with the polarization vector 
perpendicular to the plane of the disk, the atoms experience an overall dipolar repulsion and this fact 
should contribute to the stability. However, a complete numerical solution of the 
dynamics leads to the collapse
 of a strongly  disk-shaped  dipolar BEC due to the long-range 
anisotropic  dipolar interaction.
 In  the case of fermions, the stability of a trapped single-component degenerate 
dipolar Fermi gas is studied including 
the Hartree-Fock exchange and Brueckner-Goldstone correlation energies
in the local-density approximation valid for a large number of atoms.
Estimates for the maximum allowed number of polar Bose and Fermi molecules in BEC and degenerate Fermi 
gas are given.

 \end{abstract}


\maketitle

\section{Introduction}

After 
the experimental realization of Bose-Einstein condensate (BEC) of 
$^{52}$Cr \cite{pfau,4}, $^{164}$Dy \cite{7}, and $^{168}$Er \cite{8}
atoms 
and a single-component degenerate Fermi gas of  $^{161}$Dy \cite{fermi} atoms 
with large magnetic dipolar interaction,
there has been new interest  in the theoretical and experimental studies of 
degenerate gases.  
Polar  Bose \cite{bosmol} and Fermi  \cite{fermimol} molecules with much larger
permanent  electric
dipole moment  are also being  considered  for future
experiments.
The anisotropic long-range nonlocal dipolar interaction acting in all partial waves in these atoms and molecules 
is distinct from isotropic  $S$-wave contact interaction. 
Due to the anisotropic nonlocal nature of  dipolar interaction, 
the stability of a
dipolar BEC 
depends on the number of atoms, dipolar interaction and 
scattering length as well as, reasonably strongly and distinctly, on 
the trap geometry \cite{4,14,parker}. For example, in the nondipolar case the spherically symmetric 
configuration is the most stable one \cite{gammal} and accommodate the largest number of atoms, 
whereas in the dipolar case the disk-shaped configuration is 
the most stable one.

One advantage in  investigating the effect of dipolar interaction in a 
single-component
degenerate dipolar Fermi gas over that 
in a dipolar BEC is the remarkable stability of the former. Due to three-body loss via molecule 
formation, a BEC can be easily destroyed. 
The three-body loss is highly suppressed in a 
degenerate Fermi gas due to Pauli repulsion among identical fermions and this system is unconditionally 
stable in the absence of dipolar interaction. Nevertheless, in the presence of dipolar interaction beyond a 
certain strength, both the trapped BEC \cite{14,14b,parker} and  the degenerate Fermi gas \cite{zhangyi,hpu} are unstable because of possible collapse. 
New mechanism and route to collapse open up in the presence of dipolar interaction because of atomic 
interaction in non-$S$ (angular momentum $L\ne 0 $) waves. 
Dipolar BECs are immediately distinguishable from those with purely contact interaction by their strong
and distinct  
shape and stability dependence  on the trapping geometry. In the case of a trapped dipolar BEC of $^{52}$Cr atoms, anisotropic $D$ wave collapse has been studied both theoretically and experimentally \cite{18}. 
Structure formation during the collapse   has been studied in a single-component
 \cite{parker}, in a binary dipolar BEC \cite{luis}, and in a dipolar droplet bound in a nondipolar BEC
\cite{luis2}.

We undertake a systematic  study of the stability of a trapped dipolar BEC and of a 
trapped degenerate dipolar Fermi gas, both with cylindrical symmetry. 
The polarization direction is always taken along the axial symmetry direction.
Similar studies \cite{gammal} of nondipolar BECs illustrated the effect of the trapping symmetry and atomic 
contact attraction on stability.    
We present stability plots of a trapped dipolar BEC for varying number of atoms, scattering 
length, dipolar interaction, and trap symmetry. Such stability plots should aid in the future 
planning of experiments.    The repulsive atomic contact interaction as well as 
strongly 
disk-shaped trap should  facilitate stability and it is intuitively expected that these could lead to the 
absolute stability of the dipolar BEC and this possibility is 
supported by a Lagrangian variational analysis of the system. However, a complete dynamical treatment 
shows that for an increased dipolar interaction, a cylindrically trapped dipolar BEC should collapse 
independent of the underlying trap symmetry and of a large atomic contact repulsion. In the case of 
polar molecules, under the usual condition of trapping, the number of molecules in a stable BEC could 
be very small, typically, less  than a hundred. The number of molecules in a stable BEC can be increased 
by considering a significantly weaker trap with much reduced trapping frequency.

We also study, using the real-time routine, 
the dynamical nature of local collapse in dipolar BECs induced by a change of dipolar interaction, as opposed to global collapse in  nondipolar BEC induced by  a change in contact interaction managed by the Feshbach resonance 
technique \cite{fesh}. The dipolar interaction can be controlled experimentally by a rotating orienting field \cite{rotate}
to initiate the collapse while maintaining a constant atomic interaction.  The dipolar interaction is most prominent in disk  and cigar shapes and we consider only these  shapes.  In both cases we 
evidence local collapse and not a global collapse to the trap center.  In  cigar shape, the dipolar BEC is first elongated and compressed on the axial symmetry direction and with further compression, centers of collapse develop along this axis and eventually the collapsed state with small pieces of BEC 
occupies a large extension along the symmetry axis. In disk shape, due to increased dipolar repulsion,      the disk-shaped BEC first takes the shape of a donut and individual centers of collapse appear on this donut.  Eventually, after the collapse and breaking up, small pieces of 
BEC occupy the whole volume of the entire initial disk.  

The stability of a trapped degenerate dipolar Fermi gas has been studied by several groups.
 Miyakawa {\it et al.}
\cite{hpu} showed that the Fock exchange term causes a deformation of the Fermi surface in the presence of anisotropic dipolar interaction which plays a crucial role in the stability. They introduce a variational Wigner function to describe the deformation of the Fermi gas in phase space and use it to study the stability. 
Zhang and Yi \cite{zhangyi} considered the deformed Fermi surface in the form of a  ellipsoid and provided a variational and numerical solution of stability. Liu and Yin \cite{liuyin} further included a Brueckner-Goldstone (BG) correlation correction to the 
energy of a degenerate dipolar Fermi gas with deformed Fermi surface and studied its stability. 
However, we find contradiction among some of the studies and lack of physical plausibility of some of the 
results. In view of this, we find the necessity of a critical investigation of this problem.

In Sec. \ref{IIA}  we present a brief description of the mean-field model of the dipolar BEC which we use in this work. We also present an analytical variational approximation of the same. 
The present theoretical formulation for the degenerate dipolar Fermi gas in local-density approximation (LDA)
\cite{frmp}
is formulated  in Sec. \ref{IIB}. An explicit contribution of Fock deformation to Fermi surface
and of the BG correlation correction  to energy and chemical potential is included in the LDA.     We present a simple scaled form of the model which we use in numerical study. 
 In Sec. \ref{IIIA}, we present numerical results for the stability of a dipolar BEC and in Sec. \ref{IIIB}
the same for the degenerate dipolar Fermi gas.  Finally, in Sec. IV we present a 
summary of our study.

\section{Analytical Consideration}
\label{II}

\subsection{Mean-field model for dipolar BEC}

\label{IIA}
 
A dipolar BEC of $N$ atoms, each of mass $m$ satisfies the   mean-field
Gross-Pitaevskii (GP) equation \cite{pfau}
 \begin{eqnarray}  \label{gp3db} 
&& i\hbar\frac{\partial}{\partial t} {\phi{(\bf r)}}
 =  \biggr[ -\frac{\hbar^2\nabla^2 }{2m} + \frac{m \omega^2}{2}(\lambda^{-2/3}  \rho^2
 +\lambda^{4/3} z^2)\nonumber \\
&&
+\frac{4\pi \hbar^2 aN}{m} \phi^2({\bf r})+  N\int U_{\mathrm{dd}}({\bf r -r'})\phi^2({\bf r'})d{\bf r'}
\biggr] {\phi({\bf r})}, 
\end{eqnarray} 
with $\phi(\bf r)$ the wave function, 
  $a$  the atomic scattering length, $n({\bf r})
\equiv \phi^2({\bf r})$ the BEC density normalized as $\int n({\bf r}) d{\bf r}=1$,
$\lambda = \omega_z/\omega_\rho$ is the trap aspect ratio, 
$ \omega =(\omega_\rho^2 \omega_z)^{1/3}$  is the average trap angular frequency, where 
$\omega_\rho$ 
and $\omega_z$ are the angular frequencies of radial and axial traps. 
The   dipolar interaction between two atoms at $\bf r$ and $\bf r'$ 
in  (\ref{gp3db}) 
is taken as 
\begin{equation}\label{xx}
 U_{\mathrm{dd}}({\bf R}) = \frac{\mu_0\mu^2_d}{4\pi}
 \frac{(1-3\cos^2	\theta)}{R^3} \equiv  \frac{\mu_0\mu^2_d}{4\pi}   V_{\mathrm{dd}}({\bf R}) , \quad {\bf R=r-r'},
\end{equation}
 where $	\theta$ is the angle between the vector  $\bf R$ and the polarization 
direction $z$, $\mu_0$ is the permeability  of free space and $\mu_d$ is the 
magnetic dipole moment of each atom.  In case of polar molecules, each of electric dipole moment $d$, 
the prefactor in  (\ref{xx})
is $d^2/(4\pi \epsilon_0)$, where $\epsilon_0$ is the permittivity of vacuum.
To compare the strengths of atomic short-range
and dipolar interactions, the dipolar interaction is often expressed in terms of the 
length scale $a_{\mathrm{dd}}=m\mu_0\mu_d^2/(12\pi \hbar^2)$; 
for polar molecules $a_{\mathrm{dd}}=md^2/(12\pi \epsilon_0\hbar^2)$. Using this length scale, it is convenient 
to write the dipolar GP  equation (\ref{gp3db}) in the following dimensionless form
  \begin{eqnarray}  \label{3d1b} 
&&  i\hbar\frac{\partial}{\partial t} 
{\phi{(\bf r},t)}
=  \biggr[ -\frac{\nabla^2 }{2} + \frac{1}{2}(\lambda^{-2/3}\rho^2+\lambda^{4/3}z^2)
+4\pi aN \phi^2({\bf r})
\nonumber \\
&&+  3a_{\mathrm{dd}}N\int 
V_{\mathrm{dd}}({\bf r -r'})
\phi^2({\bf r'})d{\bf r'}
\biggr] {\phi ({\bf r},t)}.
\end{eqnarray} 
  In  (\ref{3d1b}) 
energy, length, density $n({\bf r})$ and time $t$ are expressed in units of 
oscillator energy
$\hbar \omega$, oscillator length 
$l_0\equiv \sqrt{\hbar/m \omega}$,  $l_0^{-3}$, and $\omega^{-1}$, respectively. 

A partial understanding of the existence of a stable trapped BEC can be obtained from a variational 
approximation to  (\ref{3d1b}) with the following ansatz \cite{lp,var}
\begin{eqnarray}\label{varan}
n({\bf r})= \frac{1}{\pi^{3/2} w_\rho^2 w_z}  \exp\left[-\frac{\rho^2}{w_\rho^2} -\frac{z^2}{w_z^2}  \right],
\end{eqnarray}
where $w_\rho$ and $w_z$ are the variational widths along radial $\rho$ and axial $z$ 
directions.
  The effective energy per particle of a stationary state propagating as 
$\phi({\bf r},t) \sim \exp({-}i\mu t/\hbar)\varphi({\bf r})$, where $\mu$ is the chemical potential,
 can be written as 
\begin{eqnarray}
&&{ E}=\int d{\bf r} \biggr[ \frac{1}{2}|\nabla_{\bf r}{\varphi(\bf r)}|^2
+\frac{1}{2}(\lambda^{-2/3}\rho^2+\lambda^{4/3} z^2)\varphi^2({\bf r})
\nonumber \\&&
+2\pi a N \varphi^4({\bf r})
+\frac{3a_{\mathrm{dd}}N}{2}\int d{\bf r'}\varphi^2({\bf r})
\varphi^2({\bf r'}) V_{\mathrm{dd}}({\bf r -r'})
\biggr].\label{efen}
\end{eqnarray}
With the density (\ref{varan}), the effective energy (\ref{efen}) becomes
\begin{eqnarray}\label{energyb}
{ E}= \frac{1}{2w_{\rho}^2}+
\frac{1}{4w_{z}^2}
+\frac{ w_{\rho}^2}{2\lambda^{2/3}}+\frac{\lambda^{4/3} w_z^2}{4}
+\frac{N[a-a_{\mathrm{dd}}f(\kappa)]   }{\sqrt{2\pi} w_{\rho}^2w_{ z}} 
 ,
\end{eqnarray}
where $ \kappa = w_\rho/w_z$ and 
\begin{eqnarray}
f(\kappa)=\frac{1+2\kappa^2}{1-\kappa^2} -\frac{3\kappa^2\mbox{tanh}^{-1}
\sqrt{1-\kappa^2}}{(1-\kappa^2)^{{3/2}}}. \label{eqn:fkappa}
\end{eqnarray}
The stable stationary states correspond to { a global minimum of energy (\ref{energyb}) for the ground state BEC.  Maxima and saddle points
 have been identified to be excited
stationary states \cite{huepe}.
The  
widths of the ground state BEC with Gaussian ansatz}
can be obtained by a minimization of energy  
(\ref{energyb}) by $\partial E/\partial w_z= \partial E/\partial 
w_\rho=0$ \cite{4,var}:
\begin{eqnarray}\label{var1}
 \frac{w_{\rho}}{\lambda^{2/3}}  &&=
\frac{1}{w_{\rho}^3} +\frac{
[a-a_{\mathrm{dd}}g(\kappa)/2]}{\sqrt{2\pi}} \frac{2N}{w_{\rho}^3w_{z}}
, \\  {w}_{z} \lambda ^{4/3} &&=
\frac{1}{w_{z}^3}+ \frac{[a- a_{\mathrm{dd}}h(\kappa)]}{\sqrt{2\pi}}
\frac{2N }{w_{\rho}^2w_{z}^2} 
,\label{var2}
\end{eqnarray}
where 
\begin{eqnarray}
&& g(\kappa) = \frac{2-7\kappa^2-4\kappa^4}{(1-\kappa^2)^2} +
\frac{9\kappa^4\mbox{tanh}^{-1}\sqrt{1-\kappa^2}}{(1-\kappa^2)^{5/2}}, \\
&& h(\kappa) = \frac{1+10\kappa^2-2\kappa^4}{(1-\kappa^2)^2} -
\frac{9\kappa^2\mbox{tanh}^{-1}\sqrt{1-\kappa^2}}{(1-\kappa^2)^{5/2}}.
\end{eqnarray}
The variational estimate for chemical potential $\mu$ is:
\begin{eqnarray}\label{mub}
\mu= \frac{1}{2w_{\rho}^2}+
\frac{1}{4w_{z}^2}
+\frac{ w_{\rho}^2}{2\lambda^{2/3}}+\frac{\lambda^{4/3} w_z^2}{4}
+2\frac{N[a-a_{\mathrm{dd}}f(\kappa)]   }{\sqrt{2\pi} w_{\rho}^2w_{ z}}.
\end{eqnarray}

\subsection{Local density approximation for polarized dipolar fermions}
\label{IIB}

Based on a hydrodynamic description of the trapped degenerate dipolar gas of 
$N$ single-component fermions each of 
mass $m$, we derived 
the following mean-field equation for this system \cite{as,dfer}
\begin{eqnarray}  \label{gp3df} 
&& \mu\sqrt{n{(\bf r)}}
 =  \biggr[ -\frac{\hbar^2\nabla^2 }{8m} + \frac{m \omega^2}{2}(\lambda^{-2/3} \rho^2 +
\lambda^{4/3} z^2)\nonumber \\
&&
+\mu_H
+  N\int U_{\mathrm{dd}}({\bf r -r'})n({\bf r'})d{\bf r'}
\biggr] \sqrt{n({\bf r})}, \\
&&\mu_H=\frac{\hbar^2}{2m}[6\pi^2 N n({\bf r})]^{2/3},
\end{eqnarray} 
and applied it \cite{dfg} to study statics and dynamics of a trapped degenerate $^{161}$Dy gas.
In  (\ref{gp3df}) $\mu$ is the chemical potential of the trapped gas, $\mu_H$ is the direct 
Hartree  
bulk chemical potential of a uniform gas with spherically symmetric Fermi surface, 
$n(\bf r)$ is the density and the dipolar 
interaction $U_{\mathrm{dd}}$ and the trapping parameters $\lambda$ and $\omega$
are the same as in the case of bosons.

The above consideration   takes the Fermi surface to be spherical
in the presence of the direct Hartree contribution to energy. However, in 
the presence of anisotropic dipolar interaction the Fermi surface will be deformed \cite{zhangyi,hpu,liuyin,bohn} and will take 
an ellipsoidal shape due to the exchange Fock interaction to energy. It has been shown that
due to this deformation of the Fermi surface, the constant Fermi 
wave vector $k_F$ for the noninteracting Fermi gas gets deformed to \cite{bohn} 
\begin{equation}
k(\theta)= k_F  +\frac{a_{\mathrm{dd}}}{3\pi}k_F^2 (3\cos^2\theta-1).
\end{equation}
This result is valid in the lowest  order in the dipolar interaction strength $a_{\mathrm{dd}}$. To this order 
this deformation  does not change the volume of the Fermi surface or the total energy or the bulk chemical potential. This leads to the following Hartree-Fock (HF) contribution to
the bulk chemical potential in the second order of interaction strength \cite{liuyin,bohn}
\begin{eqnarray}
\mu_{HF}=\mu_H
-
\frac{28\hbar^2}{135m}(6\pi^2)^{1/3}( 3a_{\mathrm{dd}}   )^2 \{Nn({\bf r})\}^{4/3}.
\end{eqnarray}

In addition to the HF contribution to energy, there is another important contribution for the 
distorted Fermi surface.  
Kohn and Luttinger \cite{kl}
showed that, in the presence of an anisotropic Fermi surface, in addition 
to the HF contribution the bulk chemical potential another contribution from the Brueckner-Goldstone (BG)
formula could be important. Lee and Yang \cite{ly} showed that  the usual BG
second-order perturbative correction $E_{BG}^{(2)}$
to the ground-state energy of a dipolar Fermi gas  has the following form:
\begin{equation}
 E_{BG}^{(2)}= \sum_{m\ne 0}\frac{\langle 0 |U_{\mathrm{dd}}|m\rangle \langle m |U_{\mathrm{dd}}|0\rangle}{E_0-E_m},
\end{equation}
where $|0\rangle$ and $|m\rangle$ are the ground and excited states of the noninteracting Fermi gas.
This term has been evaluated by Monte Carlo integration and contributes the following  to the 
bulk chemical potential \cite{liuyin}
\begin{equation}
 \mu_{BG}^{(2)}=-\frac{14}{9} \frac{\hbar^2}{m} (3a_{\mathrm{dd}})^2 \{Nn({\bf r})\}^{4/3},
\end{equation}
so that the net bulk chemical potential including the HF (exchange) and BG (correlation) correction is 
 given by
\begin{eqnarray}
\mu_{HF+BG}=
\mu_{HF}
-
\frac{14\hbar^2}{9m}( 3a_{\mathrm{dd}}   )^2 \{Nn({\bf r})\}^{4/3}.
\end{eqnarray}

The quantum pressure 
(gradient) term in  (\ref{gp3df}) contributes much less than the 
dominant ``Fermi energy" term $\hbar^2[6\pi^2Nn({\bf r})]^{2/3}/(2m)$ and its neglect 
 leads to the  LDA \cite{frmp,as,dfg}
\begin{eqnarray}  \label{gp3dTF} 
 &&\mu
 =   \frac{1}{2}m \omega^2(\lambda^{-2/3}\rho^2   +\lambda^{4/3} z^2)
+\mu_0\nonumber \\
&&+  \frac{3a_{\mathrm{dd}}N\hbar^2}{m}\int V_{\mathrm{dd}}({\bf r -r'})
n({\bf r'})d{\bf r'}, 
\end{eqnarray} 
where we have introduced a generic bulk chemical potential $\mu_0$ for the uniform gas which can be 
$\mu_H$, $\mu_{HF}$ or $\mu_{HF+BG}$, respectively, for spherical Fermi surface, including HF exchange 
correction,
or including  HF exchange and BG correlation correction.

A convenient dimensionless form of the LDA  (\ref{gp3dTF})
can be obtained in terms of  the scaling length $\bar l\equiv N^{1/6}l_0$
where $l_0=\sqrt{\hbar/m \omega}$ is the oscillator length. 
Then using the  scaled quantities 
$\bar \rho = \rho/\bar l, \bar z= z/\bar l, 
\bar {\bf r}={\bf r}/\bar l, \bar {R}={R}/\bar l,
\varepsilon_{\mathrm{dd}}\equiv 3 N^{1/6}(a_{\mathrm{dd}}/l_0)=3N^{1/3} a_{\mathrm{dd}}/\bar l, 
\bar \mu= \mu/(N^{1/3}\hbar  \omega ), 
\bar n({\bf r})= n({\bf r})\bar  l^3, $
 (\ref{gp3dTF}) can be written as 
\begin{eqnarray}  \label{gp3dS2} 
 &&\bar \mu
 =   \frac{1}{2}(\lambda^{-2/3}\bar \rho^2+\lambda^{4/3}\bar z^2)
+
\frac{1}{2}[6\pi^2 \bar  n({\bf r})]^{2/3}\nonumber \\
&&+  \varepsilon_{\mathrm{dd}}\int\frac{1-3\cos^2 \theta }
{\bar R^3}\bar n({\bf r'})d{\bf r'}-
\beta   \varepsilon_{\mathrm{dd}}   ^2 [\bar n({\bf r})]^{4/3},
\end{eqnarray}  
with  $\int d\bar { \bf r} 
\bar n({\bf r})=1$, where $\beta =0 $ for Hartree,
$\beta= 28(6\pi^2)^{1/3}/135$ for HF, and $\beta= 28(6\pi^2)^{1/3}/135+14/9$ for HF+BG.
The scaled equation (\ref{gp3dS2}) is now independent of $N$ and can be 
solved
subject to the normalization condition to determine both the chemical potential 
and the density. This $N$-independent  universal scaling (\ref{gp3dS2})
of the hydrodynamical equation 
(\ref{gp3df}) has been possible only in the LDA (\ref{gp3dTF})
after neglecting the 
quantum pressure term. 

\section{Numerical Result for Stability}

\subsection{Dipolar BEC}
\label{IIIA}

The dipolar BEC is known to be unstable to collapse \cite{yiyou} beyond a critical strength 
of dipolar interaction
independent of trap symmetry even for repulsive contact interaction 
(positive values of scattering length $a$). In this study we consider repulsive contact 
interaction only. 
In  variational  equations (\ref{var1}) and (\ref{var2})
the contribution of the nonlocal dipolar interaction is reduced to a repulsive or attractive 
contribution contained in the terms  $-a_{\mathrm{dd}}g(\kappa)$ and   $-a_{\mathrm{dd}}h(\kappa)$
similar to the contribution of the contact interaction through the scattering length $a$. In these 
equations 
the aspect ratio $\kappa=,  <  $ and $ > 1$, corresponds to a spherical, cigar, and disk shape,
respectively.  
Note that as $\kappa$ changes from 0 (extreme cigar shape) to $\infty$ 
(extreme disk shape), $g(\kappa)/2$ and $h(\kappa)$ change from 
1 to $-2$ while $\lambda$ changes from 0 to $\infty$.
The variational approximation is an oversimplification 
of the actual state of affairs in case of dipolar interaction and fails to 
predict   collapse in several cases. { Improved variational ansatz \cite{kreibich} with time-dependent coupled Gaussian functions combined with spherical harmonics
 has been suggested and used 
in this context.}
 In extreme  cigar shape the variational 
contribution of the dipolar interaction is  attractive and in extreme disk shape it is 
repulsive.  
In the disk shape one can  
argue erroneously that the system will always be stable for any number of atoms as parallel in-plane
dipoles with dipole moments aligned perpendicular to the plane always repeal and should make the 
system more stable.  The net effect of short-range atomic and long-range dipolar interactions contained 
in the square brackets of  (\ref{var1}) and (\ref{var2})  has to be attractive (negative)
beyond a critical value for the system to collapse. For a moderate  to extreme disk-shaped trap $\kappa >1$
and the functions $g(\kappa)$ and $h(\kappa)$ in  (\ref{var1})  and (\ref{var2})
are negative, and the contribution of $a_{\mathrm{dd}}$   is positive  (repulsive), thus prohibiting any collapse.  For a moderate to extreme 
cigar-shaped trap $\kappa <1$ and the functions $g(\kappa)$ and $h(\kappa)$ are positive, 
and the net contribution  of dipolar interaction   in  (\ref{var1}) and (\ref{var2}) could be
negative  (attractive), thus leading to  collapse beyond a critical value of dipolar interaction. 
In extreme  cigar shape the functions $g(\kappa)/2$ and $h(\kappa)$ are positive but less than unity in 
magnitude. The quantities in square brackets in  (\ref{var1}) and (\ref{var2}) are always positive 
(repulsive) for $a_{\mathrm{dd}}<a$. Consequently, collapse in a cigar-shaped 
dipolar BEC in variational approximation is only possible for $a_{\mathrm{dd}}>a$ and in a strongly disk-shaped BEC
it is never possible. 
Although, the dipolar BEC is unstable
to collapse 
for large enough $N$ for all values of parameters, the variational approximation can only predict collapse for 
$a_{\mathrm{dd}}>a$.   
The chemical potential and sizes predicted by the variational approximation are in good agreement 
with the numerical results for sets of parameters where stable BECs exist. 
In addition, the variational approximation may  predict stable states for parameters,  for which the 
system actually collapses.

\begin{figure}[!t]
\begin{center}
\includegraphics[width=.49\linewidth]{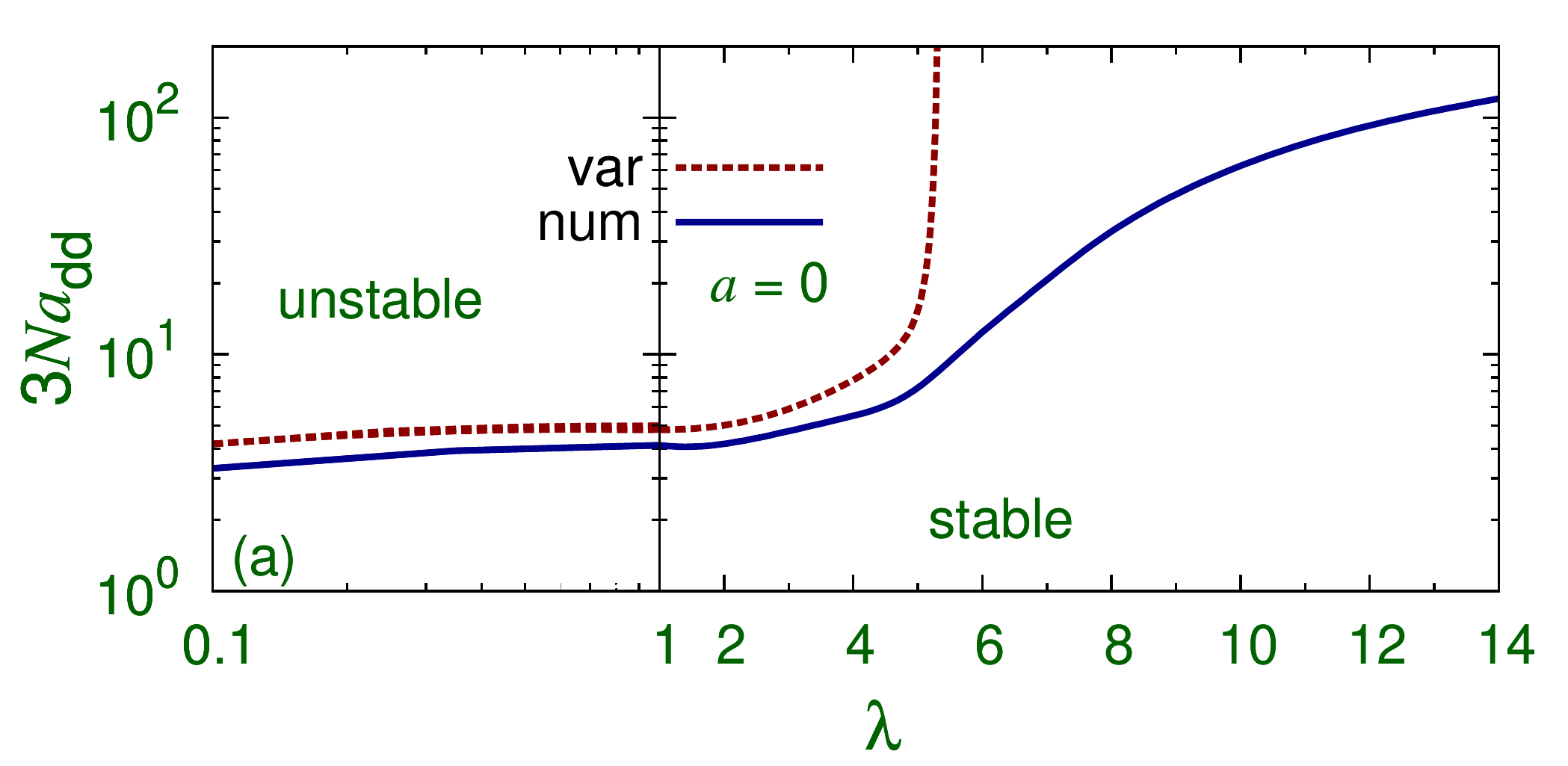}
\includegraphics[width=.49\linewidth]{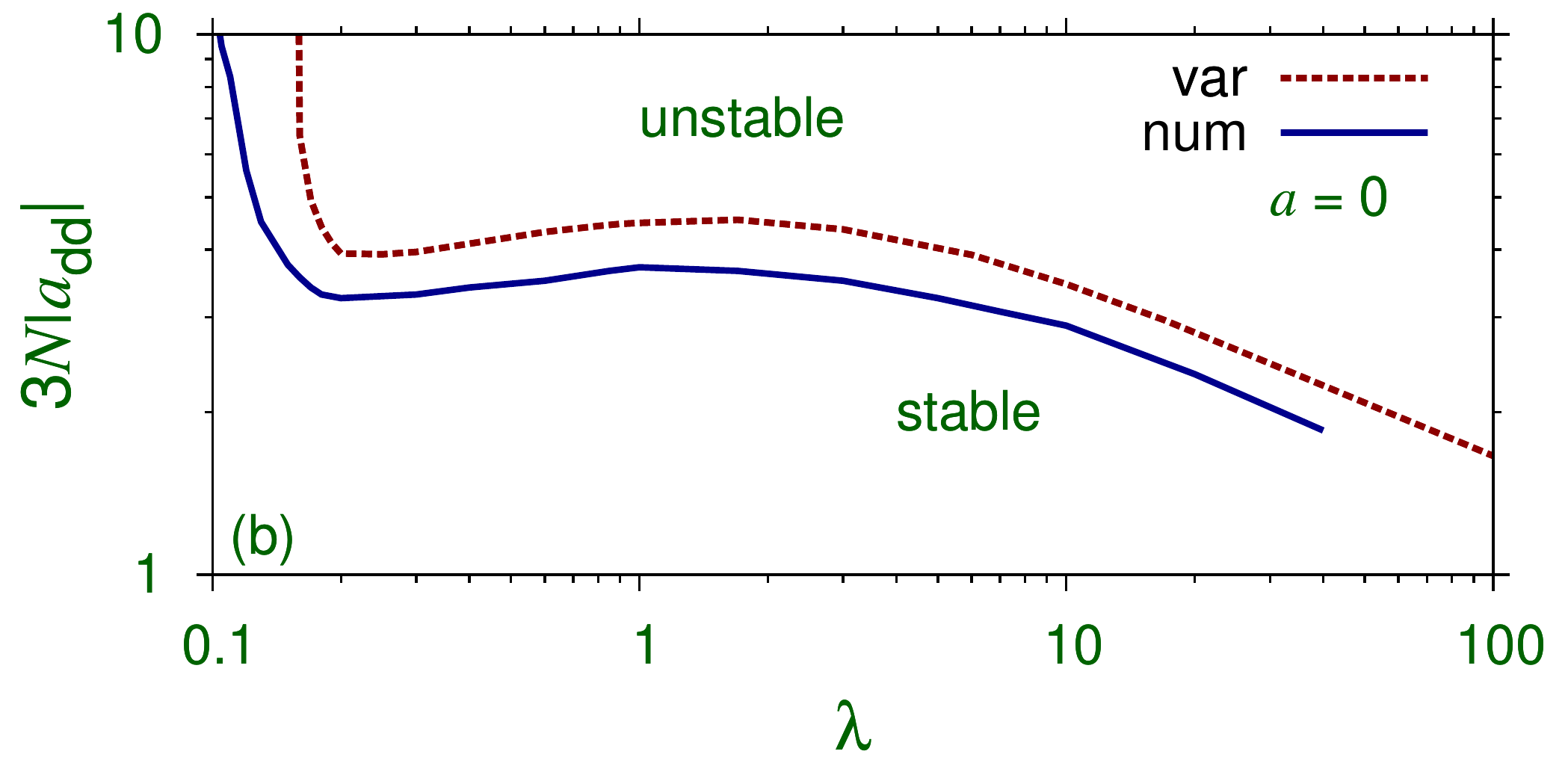}
\end{center}

\caption{  The critical value of the dipolar nonlinearity $3Na_{\mathrm{dd}}$ versus 
the trap aspect ratio $\lambda=\omega_z/\omega_\rho$ for a stable dipolar BEC in the absence of atomic 
contact interaction ($a=0$) for (a) positive and (b) negative values of the dipolar strength 
$a_{\mathrm{dd}}$ from variational approximation  (var) and numerical solution (num) of the mean-field 
equation. Length is expressed in units of $l_0=1$ $\mu$m.
}

\label{fig1}
\end{figure}

We solve the GP equation  (\ref{3d1b}) numerically by the split-step Crank-Nicolson method \cite{CPC,12}.  { The dipolar integral was evaluated in the Fourier momentum
$({\bf k})$ space using convolution theorem as \cite{12} 
\begin{equation} \label{con}
\int d{\bf r}'V_{dd}({\bf r-r}')|\phi({\bf r}')|^2
={\cal F}^{-1}\{ {\cal F}[V_{dd}]({\bf k}){\cal F}[|\phi|^2]({\bf k})\} (\bf r),
\end{equation}
where ${\cal F}[]$ and ${\cal F}^{-1}\{ \}$ are the Fourier transform (FT) and
inverse FT, respectively. The FT of the dipole potential is analytically known
\cite{12}. The FT of density $|\phi|^2$ is evaluated
numerically by means of a standard fast FT (FFT)
algorithm. The dipolar integral  (\ref{3d1b})  is
evaluated by the convolution rule (\ref{con}). The inverse
FT is taken by means of a standard FFT algorithm. The
FFT  is carried out in Cartesian coordinates
and hence the GP equation is solved in three dimensions
irrespective of the symmetry of the trapping potential.
}
We use typically a space step of 0.1 and time step 0.001 and consider up to 
512 points in space discretization in each direction. In our calculation we take the oscillator length 
$l_0 =1$ $\mu$m determined by the mass of an atom $m$ and the angular frequency $\omega$.   
{ The stability of the BEC in the numerical solution of the GP 
equation is confirmed upon obtaining a converged result after time evolution 
for a long time: typically an interval of time $\Delta t= 50$ (units of $\omega^{-1}$). If a BEC survives for such a long time, usually, it stays for 
ever and will be termed stable. }

First we consider the stability of the BEC after switching off the contact interaction by the Feshbach resonance technique ($a=0$) \cite{fesh}. This will show the effect of dipolar interaction alone. 
We consider both positive and negative values of the the dipolar strength  $a_{\mathrm{dd}}$.
The negative values of  $a_{\mathrm{dd}}$ can be achieved by a rotating orienting field \cite{rotate}.  When $a_{\mathrm{dd}}$  is negative the dipolar interaction is attractive in disk shape and repulsive in cigar shape. So, for negative $a_{\mathrm{dd}}$, the BEC is more stable in cigar shape and less in disk  shape.

 In figure \ref{fig1} (a) and (b) we present the variational and numerical stability plots for $a=0$  and for positive and negative values of $a_{\mathrm{dd}}$, respectively, for different values of trap aspect ratio $\lambda$.  The numerical results show that for all values of $\lambda$
there can be collapse for both positive and negative $a_{\mathrm{dd}}$. The variational results fail to predict collapse
for $\lambda > 5   $ in the case of positive $a_{\mathrm{dd}}$ and for $\lambda <  0.15$ in the case of negative $a_{\mathrm{dd}}$. In these cases, variational result predicts absolute stability of the dipolar BEC, whereas the numerical   result predicts a critical value of number of atoms (or of the dipolar strength $a_{\mathrm{dd}}$)
beyond which the dipolar BEC is unstable as shown in figures \ref{fig1} (a) and (b).

\begin{figure}[!t]
\begin{center}
\includegraphics[width=.49\linewidth]{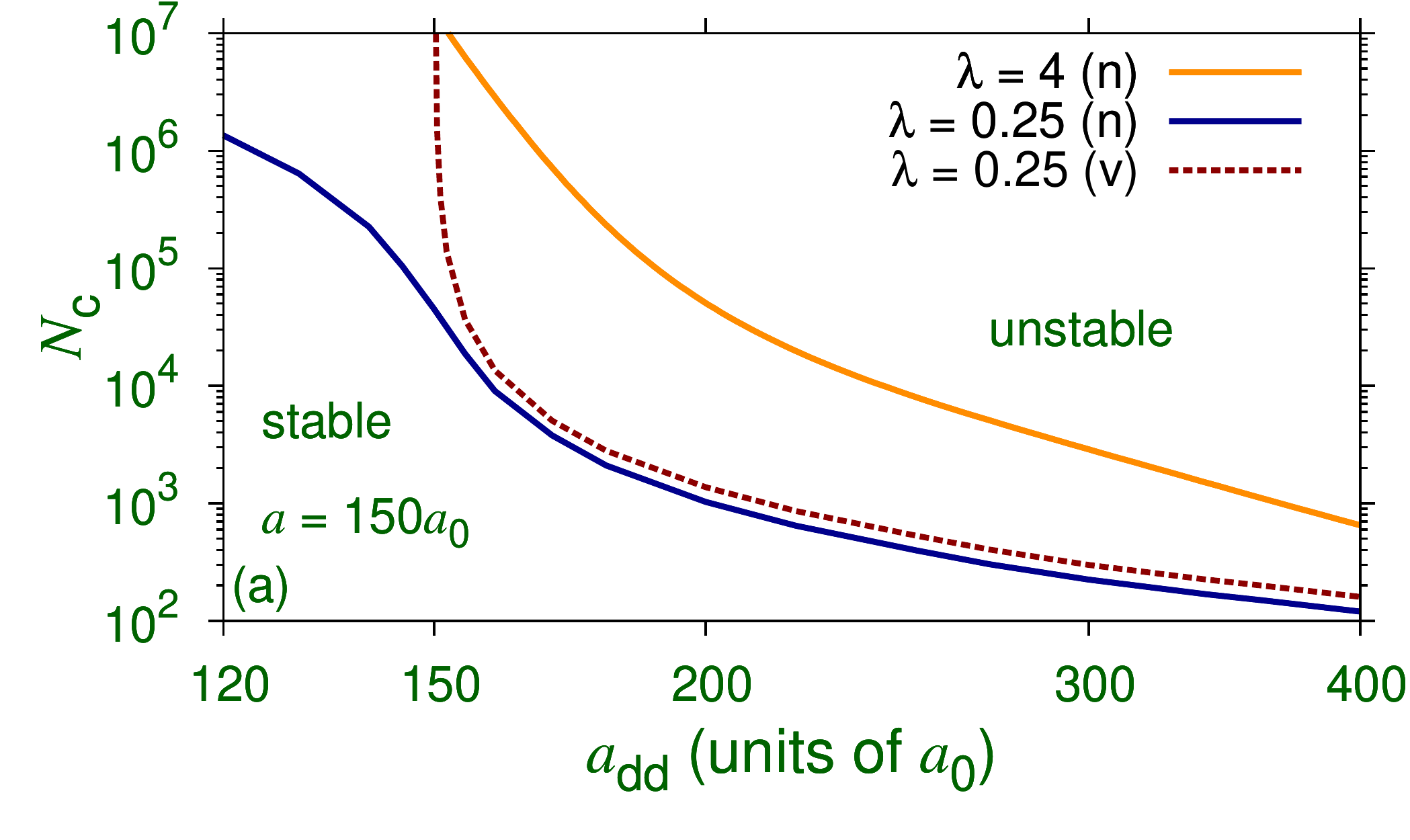} 
\includegraphics[width=.49\linewidth]{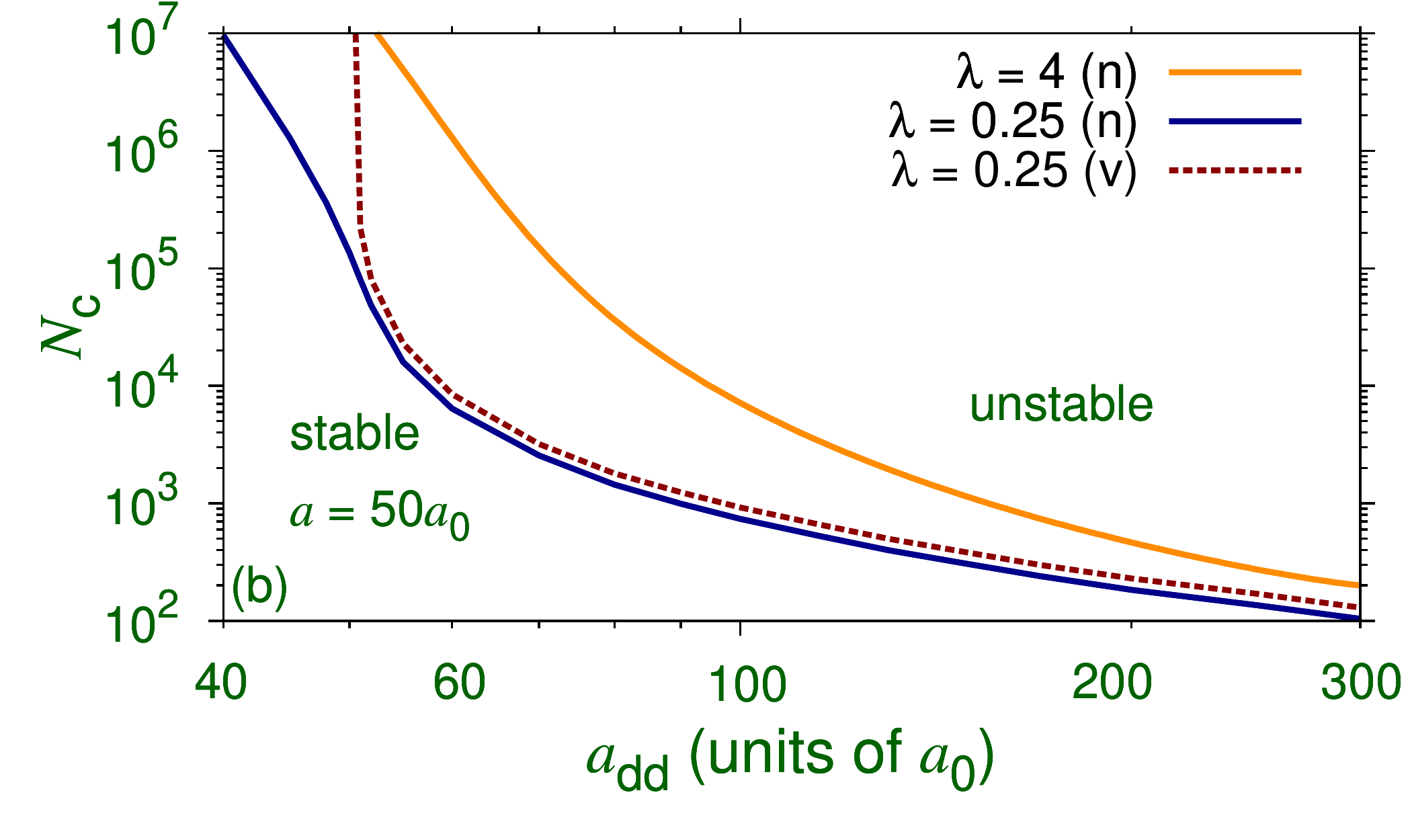}
\end{center}

\caption{The critical  number of atoms $N_c$ versus the dipolar strength 
 $a_{\mathrm{dd}}$  for a stable dipolar BEC with atomic  $S$-wave scattering length (a) $a=150a_0$
and (b)  $a=50a_0$ for trap aspect ratio $\lambda=0.25$ and 4 from variational approximation (v) and numerical solution (n) of the mean-field 
equation. Oscillator unit used in calculation is  $l_0=1$ $\mu$m.
 }

\label{fig2}
\end{figure}

\begin{figure}[!t]
\begin{center}
\includegraphics[width=.49\linewidth]{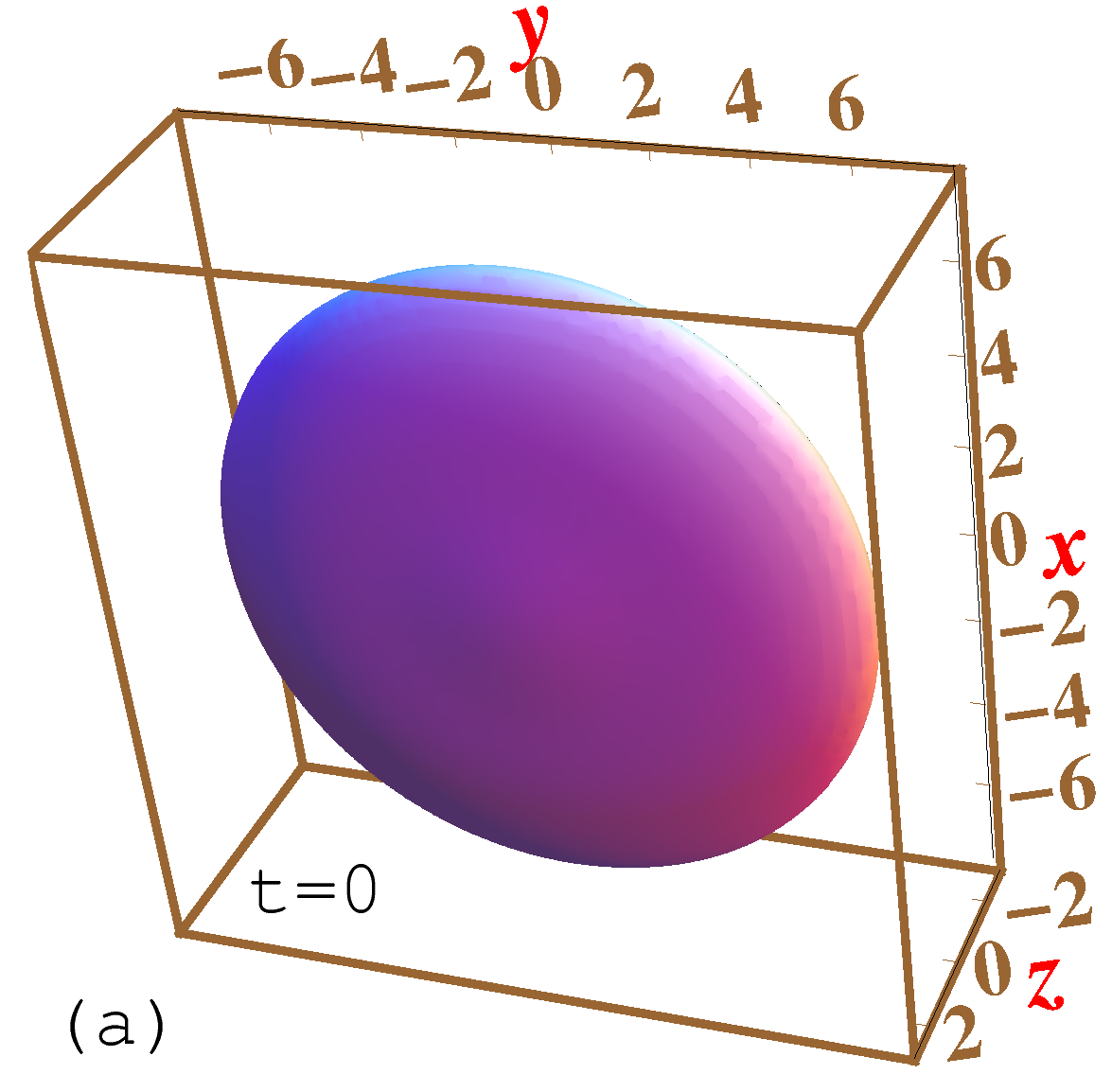}
\includegraphics[width=.49\linewidth]{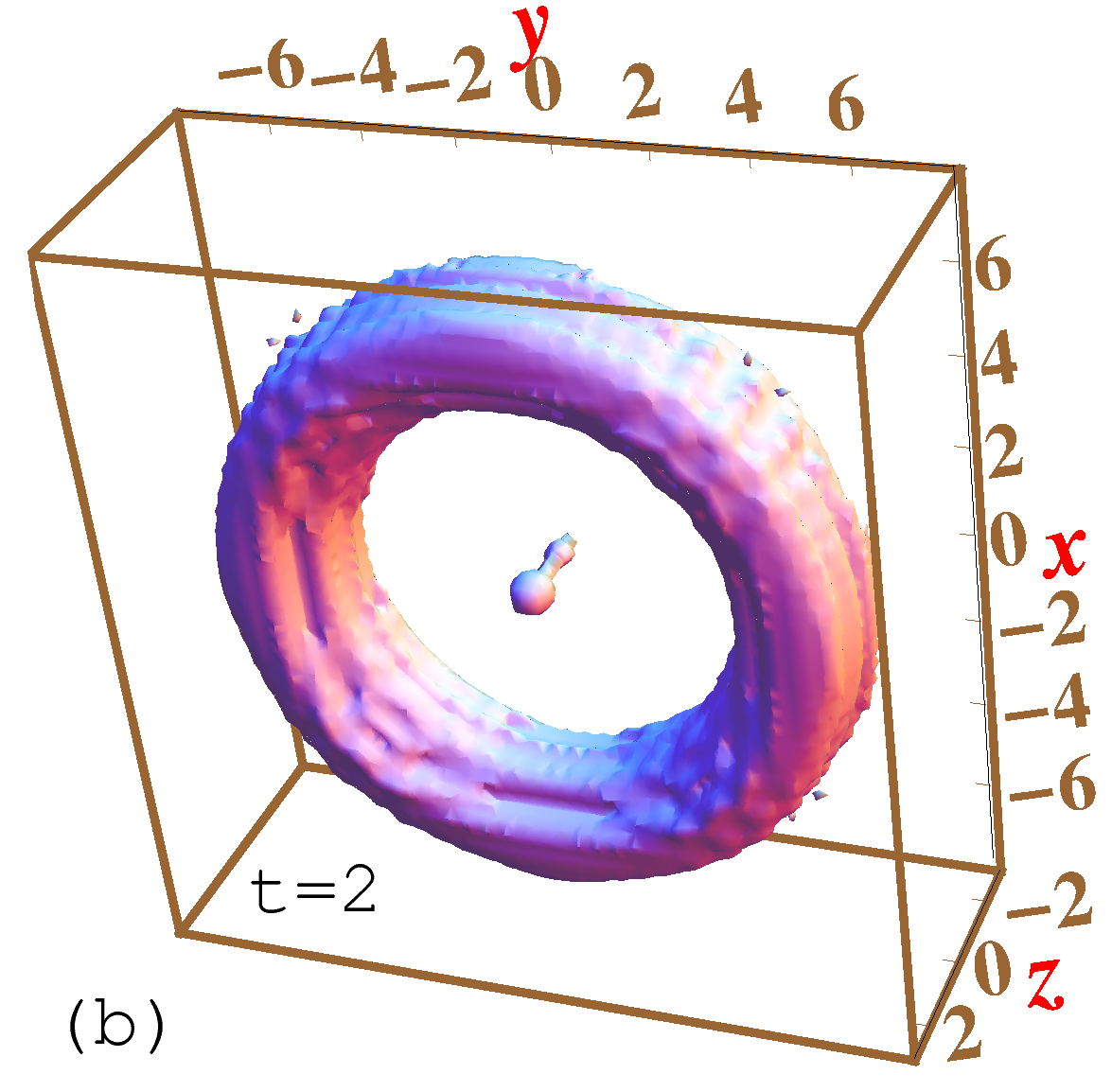}
\includegraphics[width=.49\linewidth]{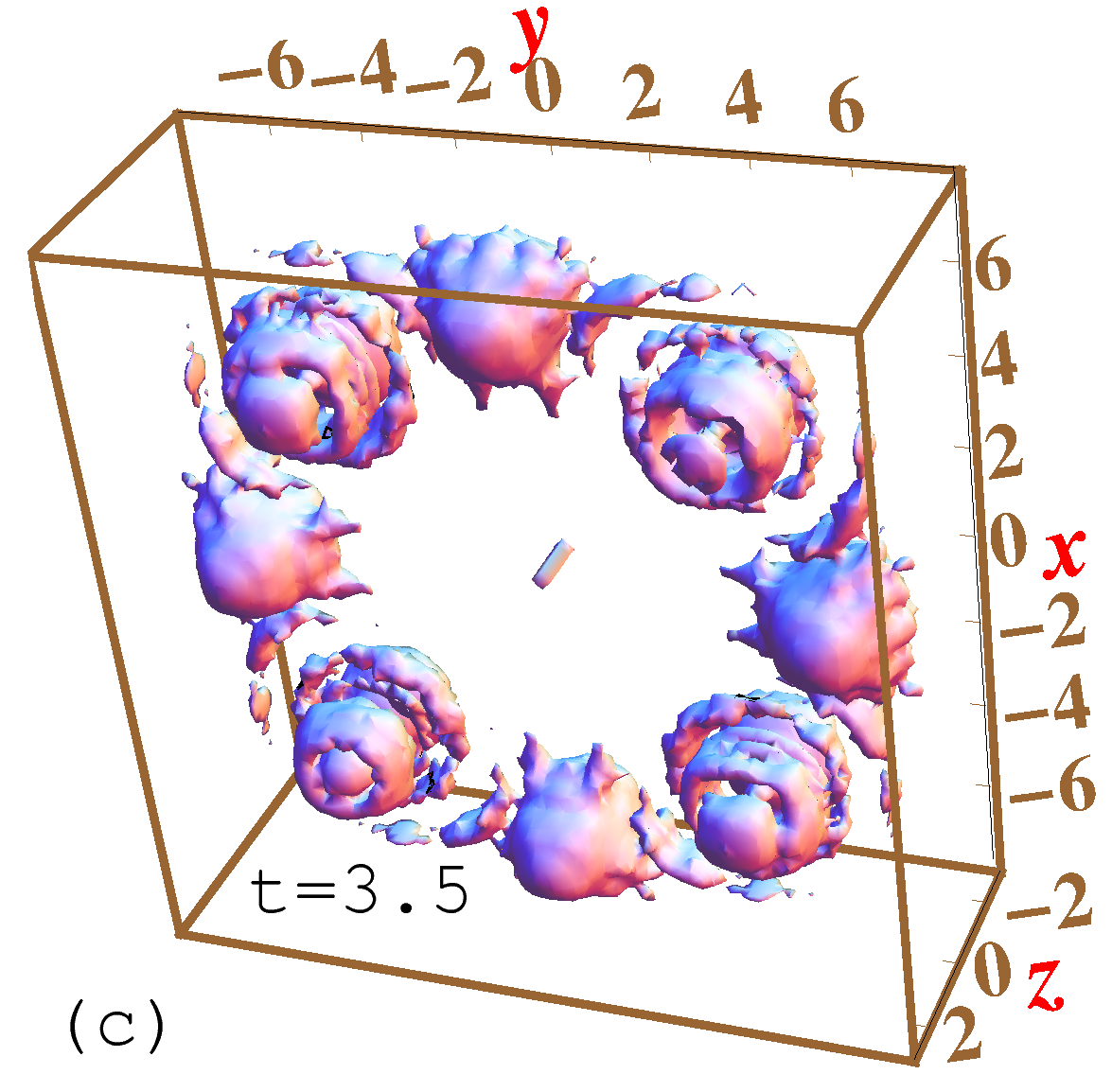}
\includegraphics[width=.49\linewidth]{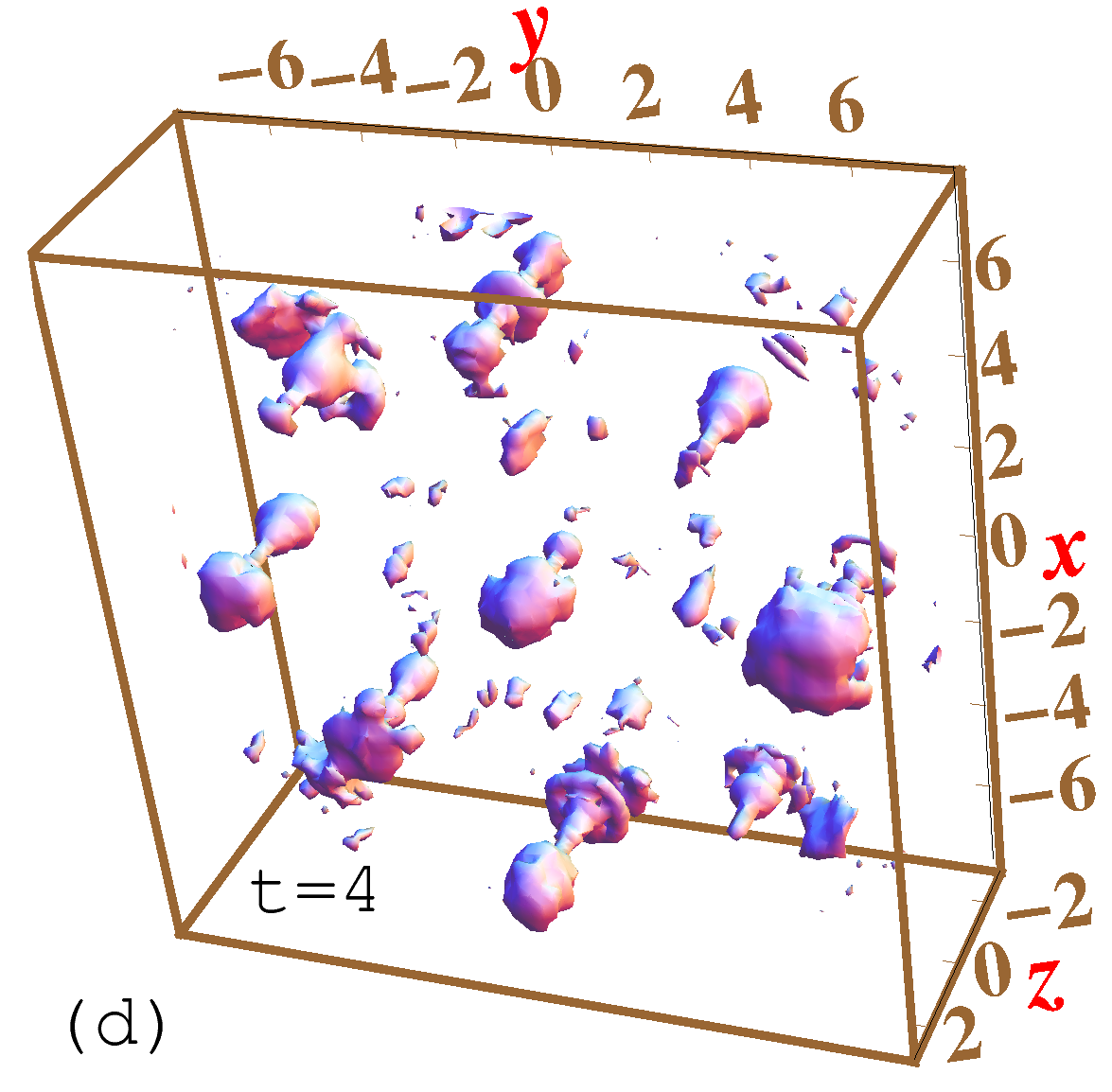}

\end{center}

\caption{Dynamics to dipolar collapse: Isodensity contour of a disk-shaped
dipolar BEC of 13000 $^{164}$Dy atoms, with scattering length $a$ set to zero by the Feshbach resonance 
technique, 
at times (a) $t=0$, (b) 2, (c) 3.5, and (d) 4. The trap parameters are $\omega=2\pi \times 60 $ Hz,
 $\lambda = 8$, $l_0= 1$ $\mu$m, $t_0= 2.65$ ms. The collapse is initiated with  
$a_{\mathrm{dd}}=16a_0$. At $t=0$ the dipolar strength $a_{\mathrm{dd}}$ is jumped to  $a_{\mathrm{dd}}=40a_0$ by a rotating orienting field. The time is expressed in units of $t_0$ and length in units of $l_0$.  Density on surface of contour is 0.01 $\mu$m$^{-3}$.
 }

\label{fig3}
\end{figure}

We also consider stability plots in the presence of a repulsive contact interaction ($a>0$) for positive $a_{\mathrm{dd}}.$
In the presence of  one more variable in the system, the scattering length $a$, we present in figure \ref{fig2} (a) the stability plot of the critical number of atoms $N_c$ versus $a_{\mathrm{dd}}$ for $a=150a_0$ for $\lambda =0.25 $ (cigar shape )
and $\lambda =4$ (disk shape), where $a_0$ is Bohr radius.  The system collapses for $N>N_c$.  The same plot for $a=50a_0$ is shown in figure 
\ref{fig2} (b). In both cases the variational result is also shown. The critical number of atoms $N_c$ decreases as $a_{\mathrm{dd}}$ increases for both disk and cigar shapes. In both cases variational result cannot predict collapse for $\lambda =4$ in the disk shape and 
predicts absolute stability for all $N$.   For the cigar shape ($\lambda =0.25$) the variational result predicts collapse for $a_{\mathrm{dd}}> a$ and the variational stability line stays very close to the numerical stability line.

{Wilson {\it et al.} \cite{14} also considered stability plot of dipolar BEC using numerical and Gaussian variational methods. Their study is, at best, complimentary to the present one. The stability plot depends on three quantities for a fixed trap aspect ratio $\lambda$: atomic scattering length $a$, dipolar strength $a_{\mathrm{dd}}$, and number of atoms $N$.  The study of Wilson {\it et al.}, more appropriate for $^{52}$Cr atoms, considers instability with a variation of scattering length $a$ for fixed $N$ and $a_{\mathrm{dd}}$. We consider instability against increase of  $N$ for different $a_{\mathrm{dd}}$. So the present study should be useful to find the maximum number of atoms in a stable dipolar BEC for different dipolar atoms and should be useful in planning future experiments. We also show the domain of applicability of the Gaussian variational method: for moderate to extreme disk shape ($\lambda >5 $) the Gaussian model always leads to stable dipolar BEC whereas the numerical analysis shows instability above a critical number of atoms. 
}

\begin{figure}[!ht]
\begin{center}
\includegraphics[width=.49\linewidth]{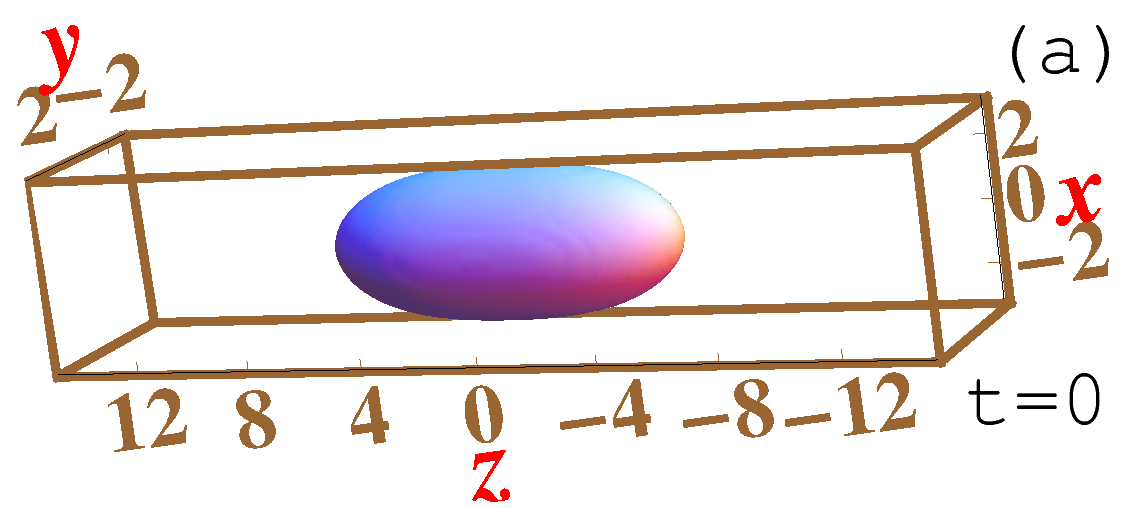}
\includegraphics[width=.49\linewidth]{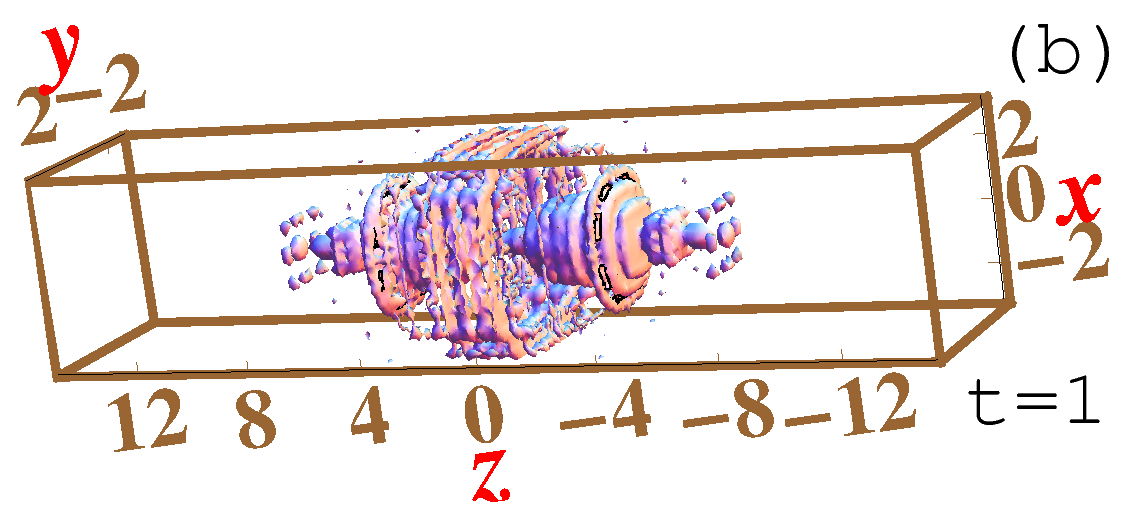}
\includegraphics[width=.49\linewidth]{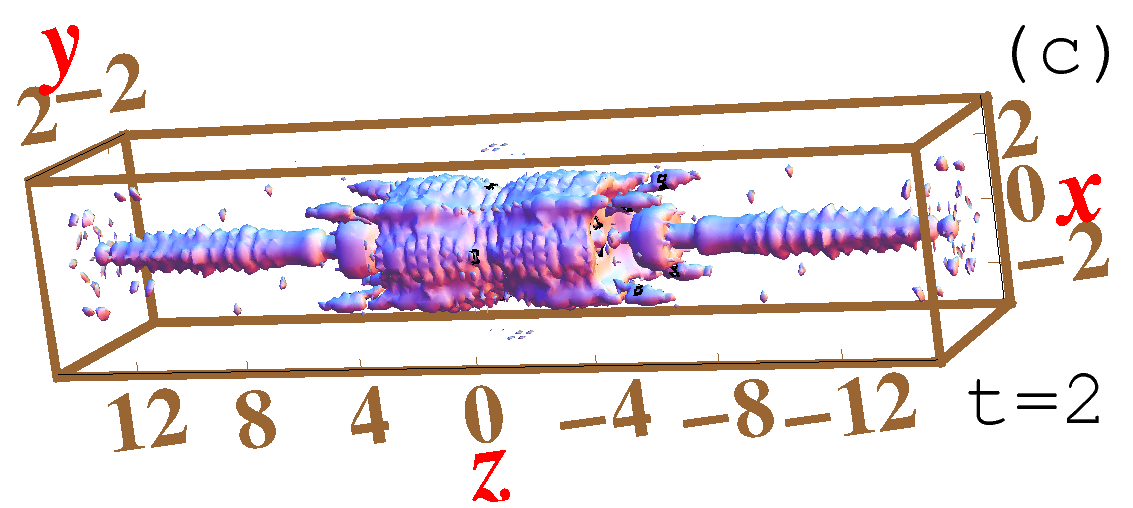}
\includegraphics[width=.49\linewidth]{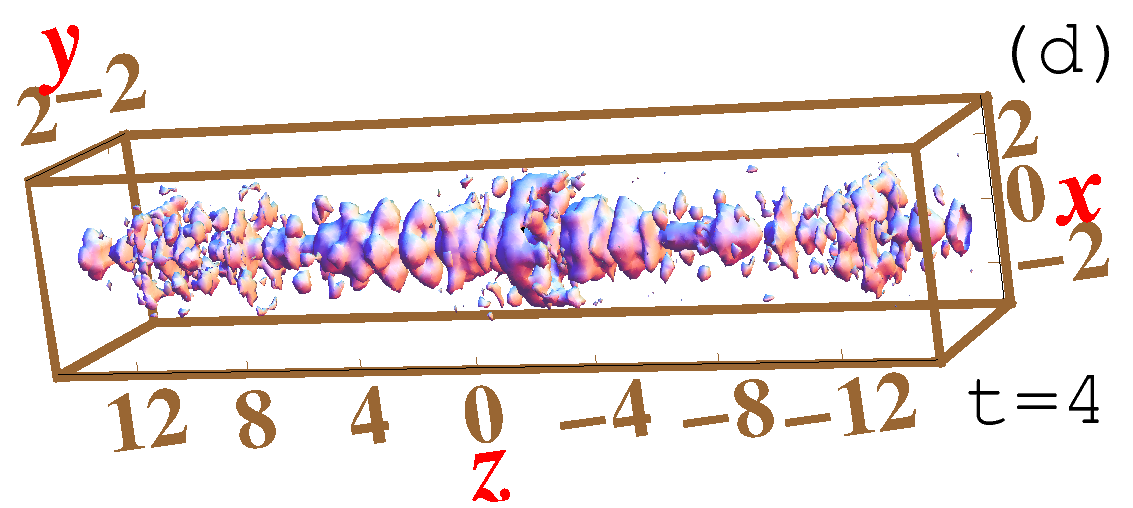}

\end{center}

\caption{  Dynamics to dipolar collapse: Isodensity contour of a cigar-shaped
dipolar BEC of 1300 $^{164}$Dy atoms, with scattering length $a$ set to zero by the Feshbach resonance 
technique, 
at times (a) $t=0$, (b) 1, (c) 2, and (d) 4. The trap parameters are $\omega=2\pi \times 60 $ Hz,
 $\lambda = 1/8$, $l_0= 1$ $\mu$m, $t_0= 2.65$ ms. The collapse is initiated with  
$a_{\mathrm{dd}}=16a_0$. At $t=0$ the dipolar strength $a_{\mathrm{dd}}$ is jumped to  $a_{\mathrm{dd}}=48a_0$ by a rotating orienting field. The time is expressed in units of $t_0$ and length in units of $l_0$.  Density on surface of contour is 0.005 $\mu$m$^{-3}$. 
 }

\label{fig4}
\end{figure}

Different experimental groups are trying to create a highly dipolar BEC of polar molecules \cite{bosmol}. The dipolar interaction strength of these molecules is typically $a_{\mathrm{dd}}=2000a_0$ \cite{pfau}. The stability lines of figures \ref{fig2}
give an idea of the maximum number of these molecules in a trap with $l_0=1$ $\mu$m. 
A preliminary calculation indicates these numbers to be only few tens.  
These numbers will increase if  
$l_0$ is increased by reducing the trapping frequency  $\omega$ as the critical number of molecules for stability is inversely proportional to $\sqrt \omega$.

Now we study the nature and dynamics of collapse in case of a disk or cigar-shaped dipolar BEC with prominent dipolar interaction by real-time evolution. In the case of spherical symmetry, the net dipolar interaction tends to average out and in that case one can have a global collapse to the center of the dipolar BEC.  To study a local collapse where a dipolar BEC collapses to different local centers we consider  disk- and cigar-shaped  dipolar BECs with $a=0$ and initiate the the collapse by jumping $a_{\mathrm{dd}}$ to a larger value by a rotating orienting 
field \cite{rotate}. In this effort, we consider $^{164}$Dy atoms with $a_{\mathrm{dd}}=130a_0$
\cite{7} and use the rotating orienting field to 
prepare an initial state with $a_{\mathrm{dd}}=16a_0$.  The trap angular frequency $\omega$ is taken as 
$2\pi \times 60$ Hz, so that the oscillator length $l_0=1$ $\mu$m. The time scale $t_0= 2.65$ ms. We consider $N=13000$ atoms in a disk shaped trap with $\lambda =8$ and initiate the collapse by jumping $a_{\mathrm{dd}}$ to 
$40a_0$ at time $t=0$ and study the subsequent real-time evolution of the dynamics. Four snapshots of the dipolar BEC at $t/t_0=0, 2, 3.5$ and 4 are shown in figures \ref{fig3}. Due to in-plane dipolar repulsion, the atoms of the 
disk-shaped dipolar BEC of figure \ref{fig3} (a)
first moves away from center and takes a donut shape as in figure \ref{fig3} (b) at $t/t_0=2$. Then due to strong out-of-plane attraction the donut-shaped BEC starts the process of local collapse and breaks up into small pieces as shown in figure 
\ref{fig3} (c) at $t/t_0=3.5$.  Finally, at $t/t_0=4$ these small pieces undergo collapse and break into smaller pieces which eventually occupy the whole spatial extension of the disk as shown in figure \ref{fig3} (d) at $t/t_0=4$.

Next we study  the local collapse 
of a cigar-shaped dipolar BEC  of $N=1300$ $^{164}$Dy atoms in a trap with 
$\omega = 2\pi \times 60$ Hz, $\lambda =1/8$, $t_0=2.65$ ms. The initial $a_{\mathrm{dd}}$ is adjusted to $16a_0$ by the 
rotating orienting field. A collapse in initiated by jumping $a_{\mathrm{dd}}$ to $48a_0$ at $t=0$.  The route 
to collapse is studied by  real-time evolution of this dynamics as shown in figures \ref{fig4}. The dipolar BEC collapses on the symmetry axis, becomes elongated and eventually breaks up into small pieces at $t/t_0=4$ along the symmetry axis. In both cases of disk and cigar shapes we have a local collapse instead of a global collapse to the global 
center of the dipolar BEC.

\begin{figure}[!t]
\begin{center}
\includegraphics[width=.49\linewidth]{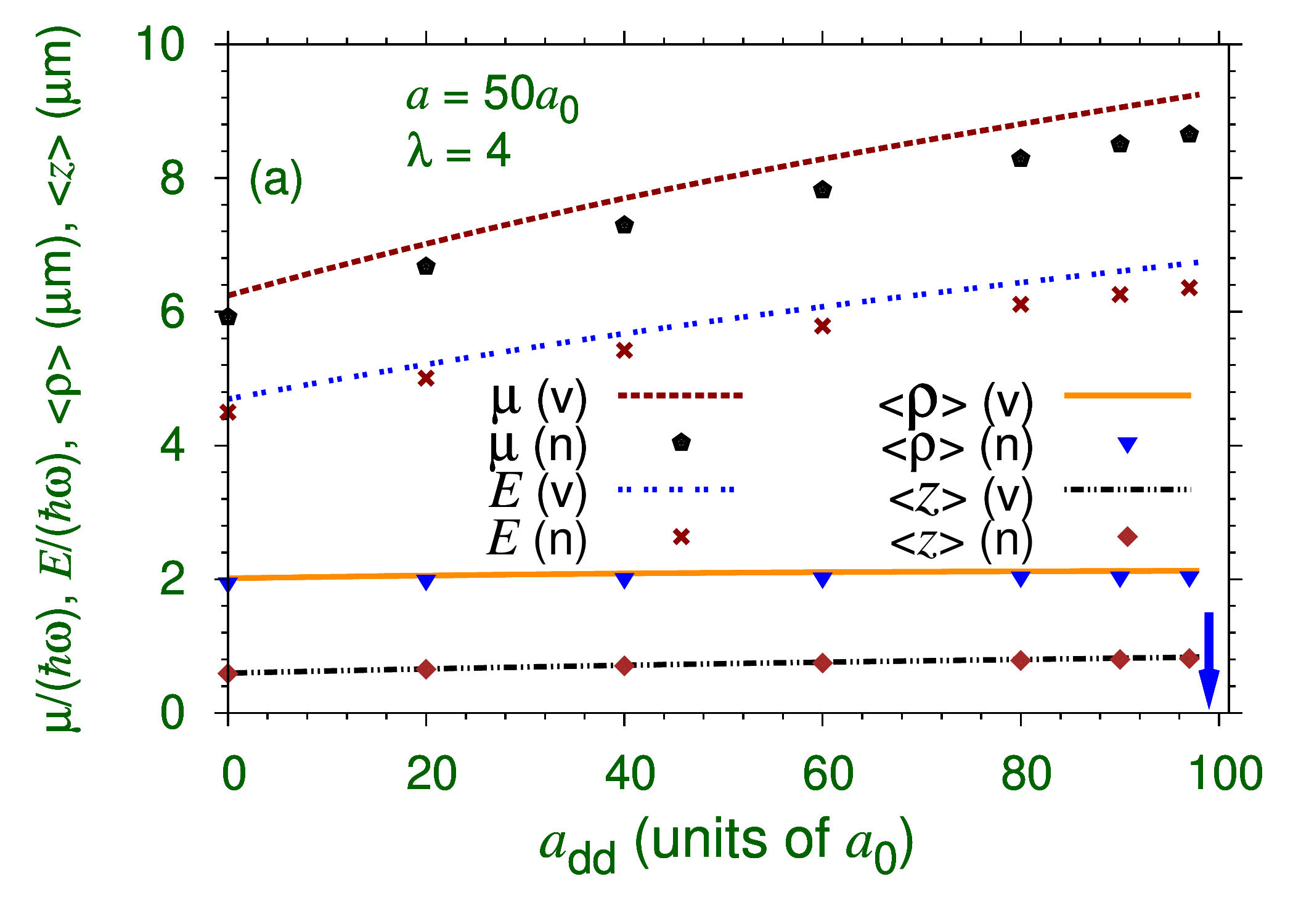}
\includegraphics[width=.49\linewidth]{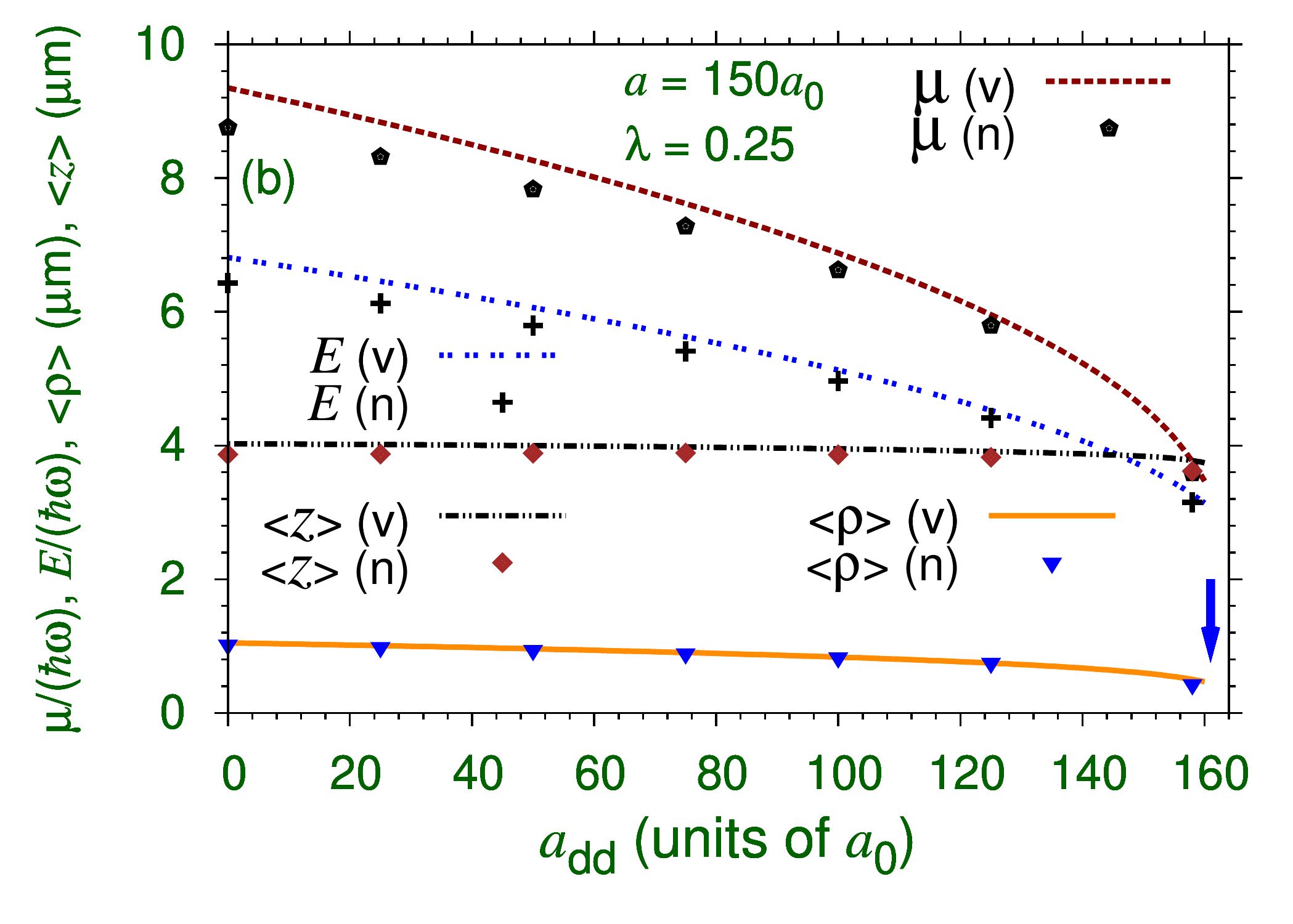}
\end{center}

\caption{ Chemical potential $\mu$, energy $E$, and rms sizes $\langle \rho \rangle$, and
$\langle z \rangle$, versus dipolar interaction strength $a_{\mathrm{dd}}$
of a dipolar BEC of 10000 atoms
 from numerical solution 
and variational approximation  of the mean-field GP equation for (a)
scattering length $a=50a_0$ and for trap aspect ratio 
$\lambda =4$ and for (b) $a=150a_0$ and $\lambda =0.25$. The oscillator length  is taken as $l_0=1$ $\mu$m.
The dark arrow on the right indicates the onset of collapse in the numerical routine at $a_{\mathrm{dd}}=
a_{\mathrm{dd}}^{(c)}$.
}

\label{fig5}
\end{figure}

{ The structures presented in figures \ref{fig3} and \ref{fig4}
do not correspond to stationary states but to nonstationary nonequilibrium 
states. Although, the trapping potential has axial symmetry, we have not 
explicitly imposed this symmetry  on the numerical solution of the GP equation. Consequently, the nonstationary nonequilibrium 
states of figures \ref{fig3} and \ref{fig4} may exhibit a small violation 
of this symmetry. It was noted before by Wilson {\it et al.} \cite{14}
that the simple Gaussian wave function assumed in  
 (\ref{varan}) cannot 
approximate the
local collapse shown, for example, in figures 
\ref{fig3} and \ref{fig4}. Furthermore, Kreibich {\it et al.} \cite{kreibich}
discussed this in detail and demonstrated
that an improved ansatz for the stability fluctuations can overcome some of
the
restrictions of the Gaussian wave function.
Also, the mean-field GP equation may not be fully adequate
for the treatment of the multi-fragmented BEC, and a full many-body description of the dynamics  might be necessary.}

We also study the chemical potential and root-mean-square (rms) sizes $\langle  \rho \rangle$ and $\langle z \rangle$
of disk- and cigar-shaped  dipolar BECs 
for parameters corresponding to stability for $0<a_{\mathrm{dd}}<a^{(c)}_{\mathrm{dd}}$ where $a^{(c)}_{\mathrm{dd}}$ is the maximum 
value of $a_{\mathrm{dd}}$ beyond which the system collapses and no stationary state can be obtained. In the disk shape the variational approximation yields   stable stationary state beyond this point, where the numerical routine 
predicts collapse.
We take $a=50a_0$ 
and $l_0=1$ $\mu$m and show the present numerical and variational results for
energy $E$, chemical potential $\mu$, and rms sizes $\langle z \rangle$ and 
 $\langle \mu \rangle$ of  
 10000 atoms in figures \ref{fig5} (a) and (b) for 
$\lambda =4 $ (disk)  and $\lambda =0.25$ (cigar), respectively.  The agreement between numerical and variational 
results is satisfactory in all cases. {The variational energies based on energy minimization are always larger than the numerical energies.}

 \subsection{Degenerate Fermi gas}

\label{IIIB}

To find the stability plot of a single-component polarized Fermi gas we use  (\ref{gp3dS2}) 
valid for a large number of fermions. We solve this equation numerically without further approximation. 
   The dipolar 
nonlinearity in  (\ref{gp3dTF}) could be very large for large $N$, whereas the allowed 
values of the scaled dipolar interaction strength $\varepsilon_{\mathrm{dd}}$ is a small number 
which makes the numerical treatment of the LDA (\ref{gp3dS2})
a relatively easy task. This inherent symmetry of the LDA of the 
dipolar Fermi gas can be turned to a good advantage in the calculation of the 
critical number of dipolar Fermi atoms in a trapped system. 
The solution of the scaled equation
 (\ref{gp3dS2}) with small nonlinearity ($N=1$) can be related to the solution 
of  (\ref{gp3dTF}) for large $N$. 
We shall see that  
(\ref{gp3dS2}) permits solution for a maximum value $\varepsilon_{\mathrm{crit}}$
of the dipolar strength
$\varepsilon_{\mathrm{dd}}$.  This can be used to find the actual critical number 
of atoms $N_c$ in a trap for a specific dipolar interaction strength  $a_{\mathrm{dd}}$ 
using the scaling relation $\varepsilon_{\mathrm{crit}} = 3N_c^{1/6}  a_{\mathrm{dd}}/l_0.$

\begin{figure}[!t]
\begin{center}
\includegraphics[width=.49\linewidth]{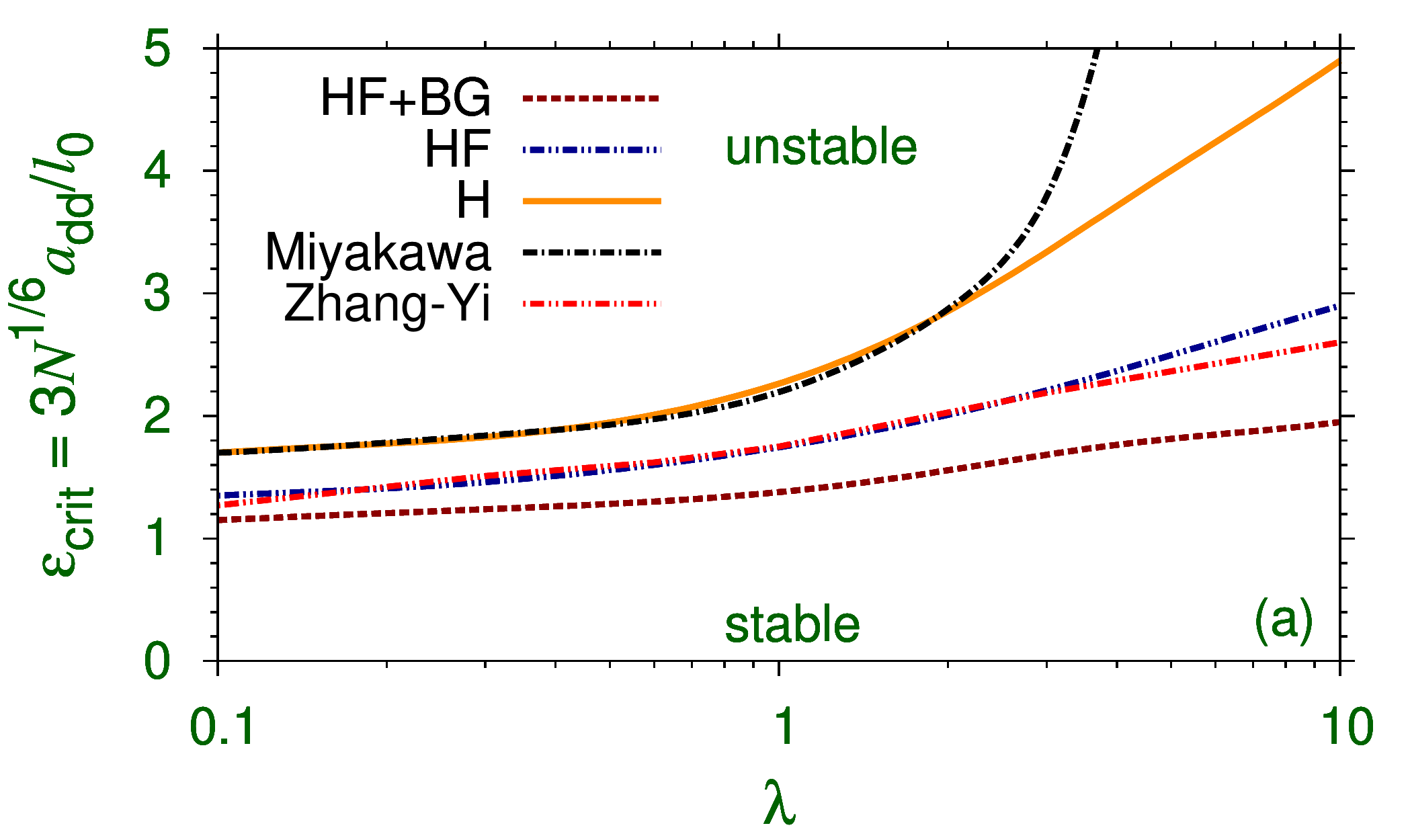}
\includegraphics[width=.49\linewidth]{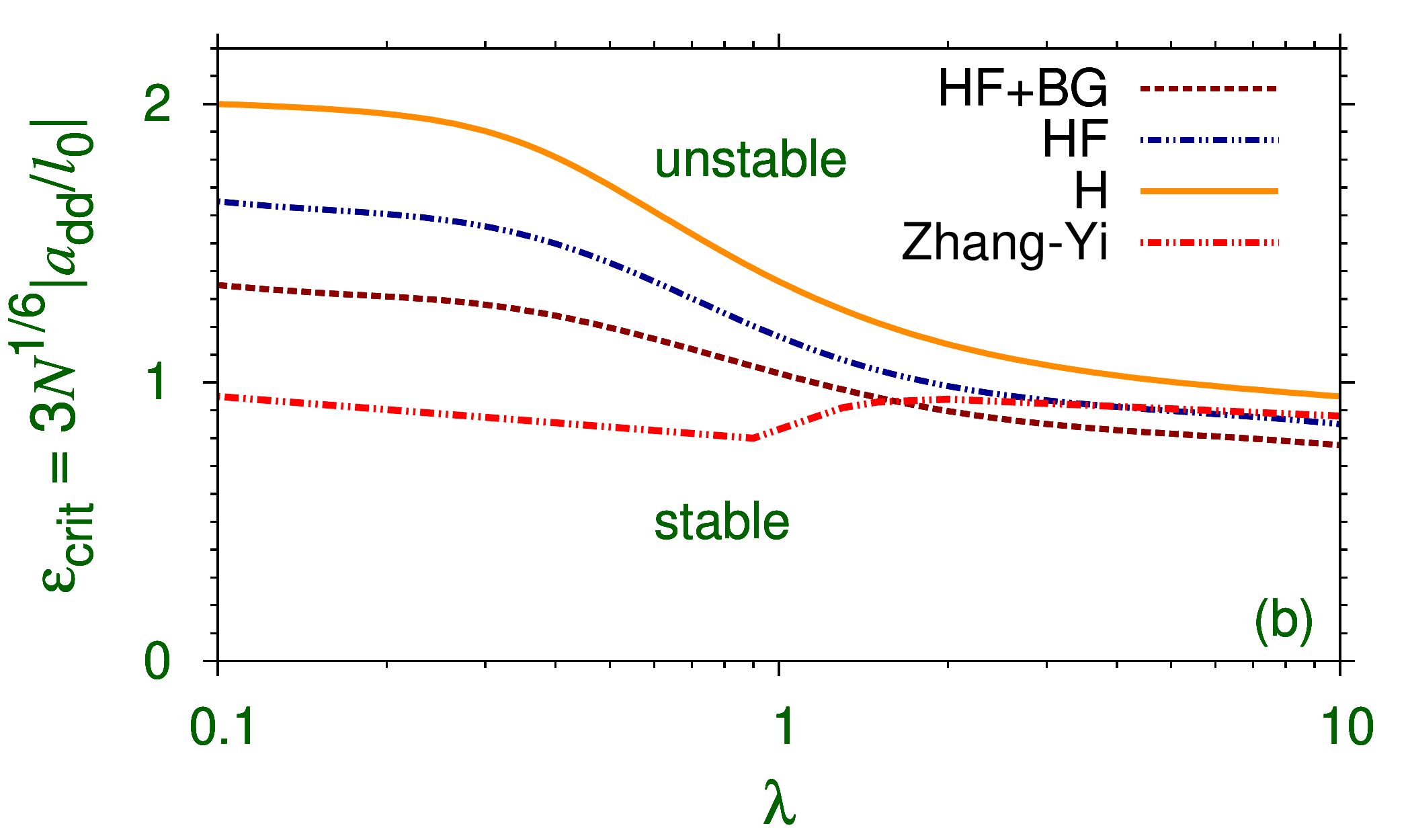}
\end{center}

\caption{The critical value of the dipolar strength   $\varepsilon_{\mathrm{crit}} \equiv 
3N^{1/6}a_{\mathrm{dd}}/l_{0}$ versus the trap aspect ratio $\lambda$ for a normal dipolar Fermi gas for 
both (a) positive and (b) negative  $a_{\mathrm{dd}}$.
}

\label{fig6}
\end{figure}

In figure \ref{fig6} (a) we plot the critical value of the scaled nonlinearity 
$\varepsilon_{\mathrm{dd}} $ versus trap 
aspect ratio $\lambda$ for positive $a_{\mathrm{dd}}$
including only the  direct Hartree  (H) contribution, 
 direct Hartree plus exchange Fock (HF) contribution, and finally the HF plus correlation BG  (HF+BG) contribution to the energy of fermions. The direct Hartree contribution to the energy is the largest, the inclusion 
of the exchange and correlation corrections reduce the energy. Hence,
the Hartree contribution  leads to the largest critical scaled nonlinearity, the 
HF and HF+BG contributions lead to smaller values of critical scaled nonlinearity as can be seen in figure \ref{fig6}
(a). In this figure we also plot the results   of Miyakawa {\it et al} \cite{hpu} and  of Zhang and Yi \cite{zhangyi}.  
{There is considerable discrepancy between these two previous
calculations obtained using similar physical models, although the approximations and calculational procedure of these two studies are not identical. In view of this we consider the contribution of the different terms in the present model and compare with these previous calculations.}
The calculation of Zhang and Yi \cite{zhangyi} including exchange  is  reasonably close to the present HF result  up to $\lambda =4$ beyond which small discrepancy is noted.  
The calculation of Miyakawa {\it et al} \cite{hpu}
including the HF contribution to energy is quite distinct. The latter
calculation 
including exchange
agrees with the present direct Hartree result, { and not with the present 
HF result,}
up to $\lambda =2$. The critical nonlinearity of      Miyakawa {\it et al} including exchange 
for $\lambda >2$ 
is much larger than the present direct Hartree result, whereas after including exchange it should be smaller. { The reason for this discrepancy 
is not fully clear, although it could be related to the approximate variational nature of this previous calculation \cite{hanpu}.}
For positive $a_{\mathrm{dd}}$ the dipolar interaction is attractive in the cigar shape and repulsive in the disk shape and consequently,  $\varepsilon_{\mathrm{crit}}$ increases as $\lambda$ changes from 0.1 to 10.  

In figure \ref{fig6} (b) we  plot the same stability lines of figure \ref{fig6} (a)
for negative $a_{\mathrm{dd}}$.  
In this figure we also show the stability line of Zhang and Yi \cite{zhangyi} including exchange.  { Considerable discrepancy is noted 
between the present HF model calculation and that of Ref.  \cite{zhangyi}
for $\lambda<2$. } For negative $a_{\mathrm{dd}}$ the dipolar interaction is attractive in the disk shape and repulsive in the cigar shape and consequently, critical $\varepsilon$ decreases as $\lambda$ changes from 0.1 to 10. However, in the stability plot  of Zhang and Yi the critical values of $\varepsilon$ for $\lambda=0.1$ and 10 are practically the same,  which does not seem physically very plausible. In view of these crucial discrepancies, 
further studies are needed to unveil the truth. 

The stability lines of figure \ref{fig6} (a) give an idea of the maximum number of polar Fermi molecules in a 
typical trap of oscillator length $l_0=1$ $\mu$m.  
Using the present HF+BG result, we can estimate the critical number of molecules for
$\lambda =0.1$ and 10 using the critical values $\varepsilon_{\mathrm{crit}}
=1.1$ and 2.5, respectively.   Using a  dipolar strength $a_{\mathrm{dd}}=2000a_0$ \cite{pfau}, we obtain the 
critical number of molecules $N_c\approx 1700$ and 240,000, respectively, for $\lambda =0.1$ and 10.  
As expected, these numbers are much larger than those in the case of bosonic polar molecules. The higher stability in case of the fermions is due to the Pauli exclusion principle among identical fermions.  

\section{Summary and Conclusion}
  
Trapped  nondipolar BECs with  repulsive atomic interaction as well as trapped degenerate single-component Fermi gas are unconditionally stable. The BECs are subject to collapse instability for atomic attraction above a critical value \cite{gammal}. Trapped dipolar BECs with a sufficiently strong dipolar interaction 
are more fragile and could be unstable to collapse independent of the underlying trap symmetry
even for repulsive contact interaction \cite{14,14b}.  The condition of stability of the dipolar BECs is fully distinct. 
We studied the conditions of stability of dipolar BECs in axially-symmetric trap for varying trap symmetry and 
varying strengths of contact and dipolar atomic interactions.  In all cases collapse is possible for a sufficiently large dipolar interaction. We demonstrate the local nature of dipolar collapse by real-time simulation of the collapse dynamics, where centers 
of local collapse develop all over the condensate and small fragments of the condensate collapse to these local centers, as opposed to a global center.  We estimate the maximum number of bosonic polar molecules in a stable condensate. For a  dipolar  interaction strength $a_{dd}=2000a_0$, the maximum number of molecules in a trap of oscillator length $l_0=1$ $\mu$m is typically few tens and this number can be increased with a weaker trap. 

We also studied the stability of a degenerate Fermi gas in stability plots. Independent of the underlying trap symmetry, this system will collapse for a sufficiently strong dipolar interaction \cite{hpu}. The maximum number of Fermi atoms allowed in a stable system is usually much larger than that permitted in a dipolar BEC under similar conditions. We also estimate  the maximum number of fermionic polar molecules in a stable system. For a typical dipolar  interaction strength $a_{dd}=2000a_0$, the maximum number of fermionic molecules in a trap of oscillator length $l_0=1$ $\mu$m is  larger than the maximum number of bosonic molecules by several orders of magnitude.

\ack

We thank FAPESP  and  CNPq (Brazil)  for partial  support.

\end{document}